\documentclass[prl,letterpaper,aps,10pt,superscriptaddress,twocolumn,floatfix,showpacs]{revtex4-1}
\usepackage{graphicx}
\usepackage{amsmath}
\usepackage{amsfonts}
\usepackage{amssymb}
\usepackage{epsfig}
\usepackage{epstopdf}
\DeclareGraphicsExtensions{.pdf,.eps,.png,.jpg,.mps}
\usepackage[pdftex]{color}
\usepackage{amsmath,graphicx,amssymb,braket,xcolor,subfigure,upgreek}
\usepackage[colorlinks, linkcolor=blue, citecolor=blue, urlcolor=blue, breaklinks=true]{hyperref}
\usepackage{microtype}
\usepackage{bbm}
\usepackage{color}

\begin{document}

\title{Laser refrigeration using exciplex resonances in gas filled hollow-core fibres}
\author{Christian Sommer}
\affiliation{Max Planck Institute for the Science of Light, Staudtstra{\ss}e 2,
D-91058 Erlangen, Germany}
\author{Nicolas Y. Joly}
\affiliation{Department of Physics, University of Erlangen-Nuremberg, Staudtstra{\ss}e 2,
D-91058 Erlangen, Germany}
\affiliation{Max Planck Institute for the Science of Light, Staudtstra{\ss}e 2,
D-91058 Erlangen, Germany}
\author{Helmut Ritsch}
\affiliation{Institut f\"ur Theoretische Physik, Universit\"at Innsbruck, Technikerstr. 21a, A-6020 Innsbruck, Austria}
\author{Claudiu Genes}
\affiliation{Max Planck Institute for the Science of Light, Staudtstra{\ss}e 2,
D-91058 Erlangen, Germany}
\date{\today}

\begin{abstract}
We theoretically study prospects and limitations of a new route towards macroscopic scale laser refrigeration based on exciplex-mediated frequency up-conversion in gas filled hollow-core fibres. Using proven quantum optical rate equations we model the dynamics of a dopant-buffer gas mixture filling an optically pumped waveguide. In the particular example of alkali-noble gas mixtures, recent high pressure gas cell setup experiments have shown that efficient kinetic energy extraction cycles appear via the creation of transient exciplex excited electronic bound states. The cooling cycle consists of absorption of lower energy laser photons during collisions followed by blue-shifted spontaneous emission on the atomic line of the alkali atoms. For any arbitrary dopant-buffer gas mixture, we derive scaling laws for cooling power, cooling rates and temperature drops with varying input laser power, dopant and buffer gas concentration, fibre geometry and particularities of the exciplex ground and excited state potential landscapes.
\end{abstract}


\maketitle
Lasers are  sources of energy which is highly concentrated in space and momentum at almost zero entropy~\cite{boukobza2013breaking}. This effective very low temperature is very successfully used to cool dilute atomic gases to almost arbitrarily close to absolute zero. In addition laser light generated optical traps can be considered to have zero temperature walls~\cite{phillips1998nobel}. Similarly, opto-mechanical cooling of single eigenmodes of microscopic mechanical oscillators can reach the quantum ground state~\cite{chan2011laser,genes2009micromechanical}. However, despite long standing efforts, applications towards cooling molecular gases~\cite{zhelyazkova2014laser,shuman2010laser,hemmerling2016laser}, liquids or in particular whole solid objects~\cite{rayner2003condensed,sheik2009laser,seletskiy2010laser}, proved significantly more difficult to implement. Already two decades ago the first basic proof-of-concept implementations used selective doping to cool a whole optical fibre \cite{mungan1997laser,rayner2001laser} by tens of degrees. Anti-Stokes Raman light scattering of the guided field modes was used to strongly depopulate phononic or motional modes~\cite{pringsheim1929zwei}. Unfortunately, non-radiative emission processes and reabsorption of the anti-Stokes photons by impurities, as well as heating from the environment and support structures limit the efficiency of this process. In the following generations of experiments, the use of isotope purified small crystals and improved environmental shielding eventually allowed to reach cryogenic temperatures of macroscopic objects~\cite{melgaard2016solid}, significantly beating the temperature limits achieved via thermo-electric cooling~\cite{disalvo1999thermoelectric}.
Despite these significant improvements over the last two decades, the efficiency of the cooling process remained rather low, reaching only a couple of miliwatts of cooling power from $50$~W of laser power. As one central reason for this low efficiency one can identify the rather limited Stokes shift of the emitted phonons of less than $1/1000$ of their frequency. Hence one needs thousands of successful cycles per background absorption with non-radiative decay. Interestingly it has been shown recently that this ratio can be tremendously improved using excited molecular states (exciplexes) of Alkali-rare gas molecules as e.g. Rb-Ar or K-Ar~\cite{Sass2011laser}. In this case the incoming laser light can be up-converted by a few percent allowing to increase the efficiency per scattering event by several orders of magnitude. In several seminal recent experiments laser cooling of high pressure (order of a hundred bar) alkali-rare gas mixtures has been experimentally demonstrated, potentially reaching much lower temperatures~\cite{Weitz2009laser,Sass2011laser,Gelbwaser2015laser}. Recently, also cooling of doped levitated particles in vacuum showed astonishingly efficient cooling of the whole particle in a dipole trap~\cite{millen2015cavity}.
In a more general sense this can be seen as a sort of reverse operation of excimer laser gain transition, where light at the excimer transition is absorbed and re-emitted at much higher frequencies~\cite{cerullo2001gas}. This opens a potentially very large range of possible implementations of such cooling schemes based on excimer laser proven gas mixtures.

\begin{figure}[b]
\includegraphics[width=0.7\columnwidth]{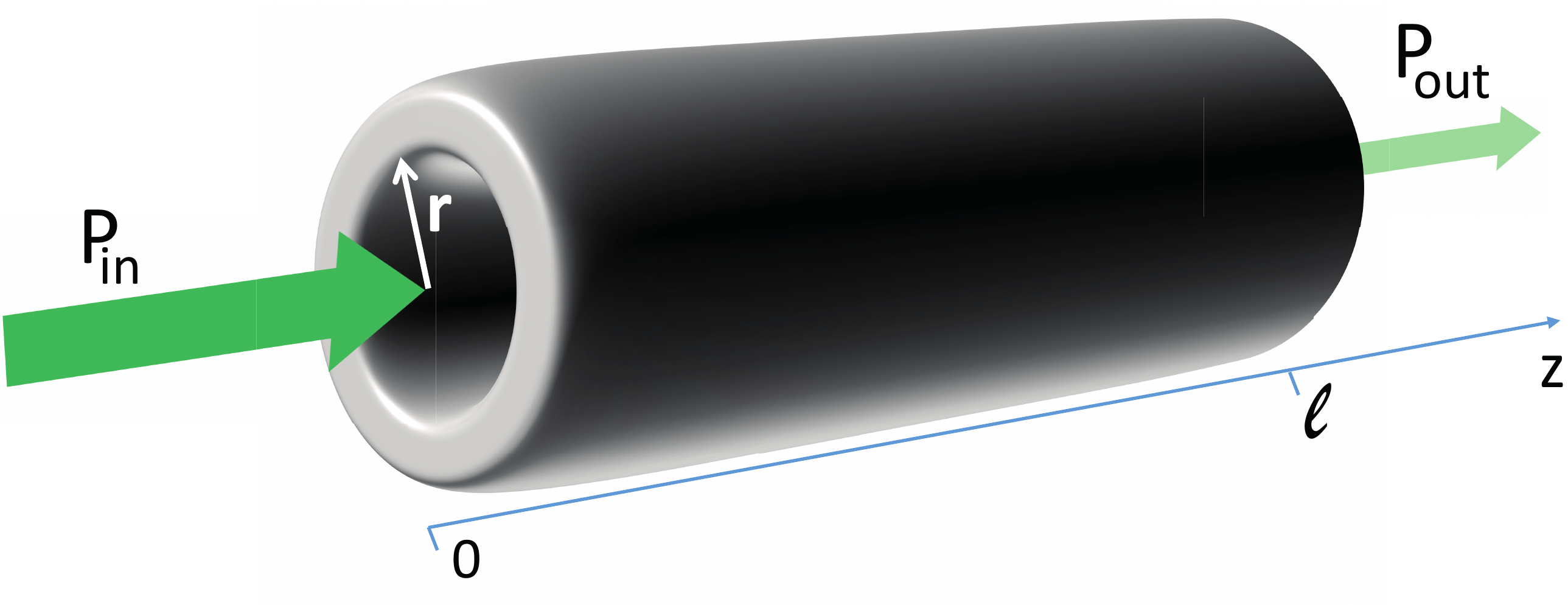}
\caption{\emph{Simplified schematics}. Hollow-core fibre of inner radius $r$ and length $\ell$ filled with M-X (dopant-buffer) gas is pumped by a laser at central frequency $\omega_{\text{L}}$ with bandwidth $\delta \omega$ and input power $\mathcal{P}_{in}$. Power loss (resulting in a reduction of the kinetic energy of the trapped gas) occurs via absorption and subsequent spontaneous emission at higher frequencies from the M atoms into unguided modes, leading to a reduced output power $\mathcal{P}_{out}$.}
\label{fig1}
\end{figure}

\noindent Laser cooling using high-pressure noble gases containing a small fraction of alkali atoms has been recently experimentally investigated~\cite{Weitz2009laser,Sass2011laser,Gelbwaser2015laser}. The mechanism is based on the particular property of exciplexes (formed by dopant, alkali atoms-M colliding with buffer, noble gas atoms-X) which exhibit a transitory bound excited state. During the picosecond time window of exciplex formation, photons from a laser (of energy $\hbar \omega_{\text{L}}$) energetically matched around the instantaneous binding energy can be absorbed. In the subsequent dynamics, the exciplex becomes unbound and up-converted blue-shifted photons at an energy $\hbar \omega_0$, which corresponds to the bare M transition can be spontaneously emitted. An overall energy loss occurs at $\hbar \Omega$ leading to a reduction of the kinetic energy of the gas and correspondingly a drop in the gas temperature. The process is described by the parameter $\hbar \Omega/(k_{\text{B}} T)$ which can reach very large values compared to similar cooling schemes.\\
\noindent Proof-of-principle experiments have shown the signature of collision-aided cooling~\cite{Weitz2009laser} in high-temperature gas cells leading to changes in the local index of refraction in the gas mixture. We aim instead at developing a general theory of refrigeration of a dopant-buffer gas mixture applicable to room-temperature environments. To this end we propose the use of highly confined guided electromagnetic modes of a gas-filled hollow-core fibre. This allows a more efficient light-gas interaction as well as the subsequent refrigeration of the gas and of the surrounding fibre walls. Our analytical model unravels scaling laws of the cooling power, cooling rate and temperature drop as a function of fibre dimensions, density of M and X gases, input power and initial temperature of the fibre glass walls. For realistic parameters, based on the particular Rb-Ar exciplex choice, we find that the cooling efficiency can surpass values achieved in state-of-the-art solid state cooling schemes reaching more than $1\%$ conversion of input power into useful cooling power. At low dopant density and reduced input power we find that higher cooling efficiencies can be achieved by an increase in buffer gas pressure as a means of enhancing the collision rate. In such a case the power absorption along the fibre follows a standard Beer-Lambert exponential reduction law. However, this tuning knob does not always provide the optimal strategy as the opposite regime of dense dopant (or higher input power) leads to a linear power absorption law: in such a case an increase in buffer gas density can actually lead to reduced cooling efficiency. Under room temperature conditions, we show that particular arrangements of laser driven gas-filled fibres in bundles can lead to a considerable temperature gradient towards the center of the structure even when only minimal temperature drops are achievable for single fibres.\\

\begin{figure}[t]
\includegraphics[width=0.98\columnwidth]{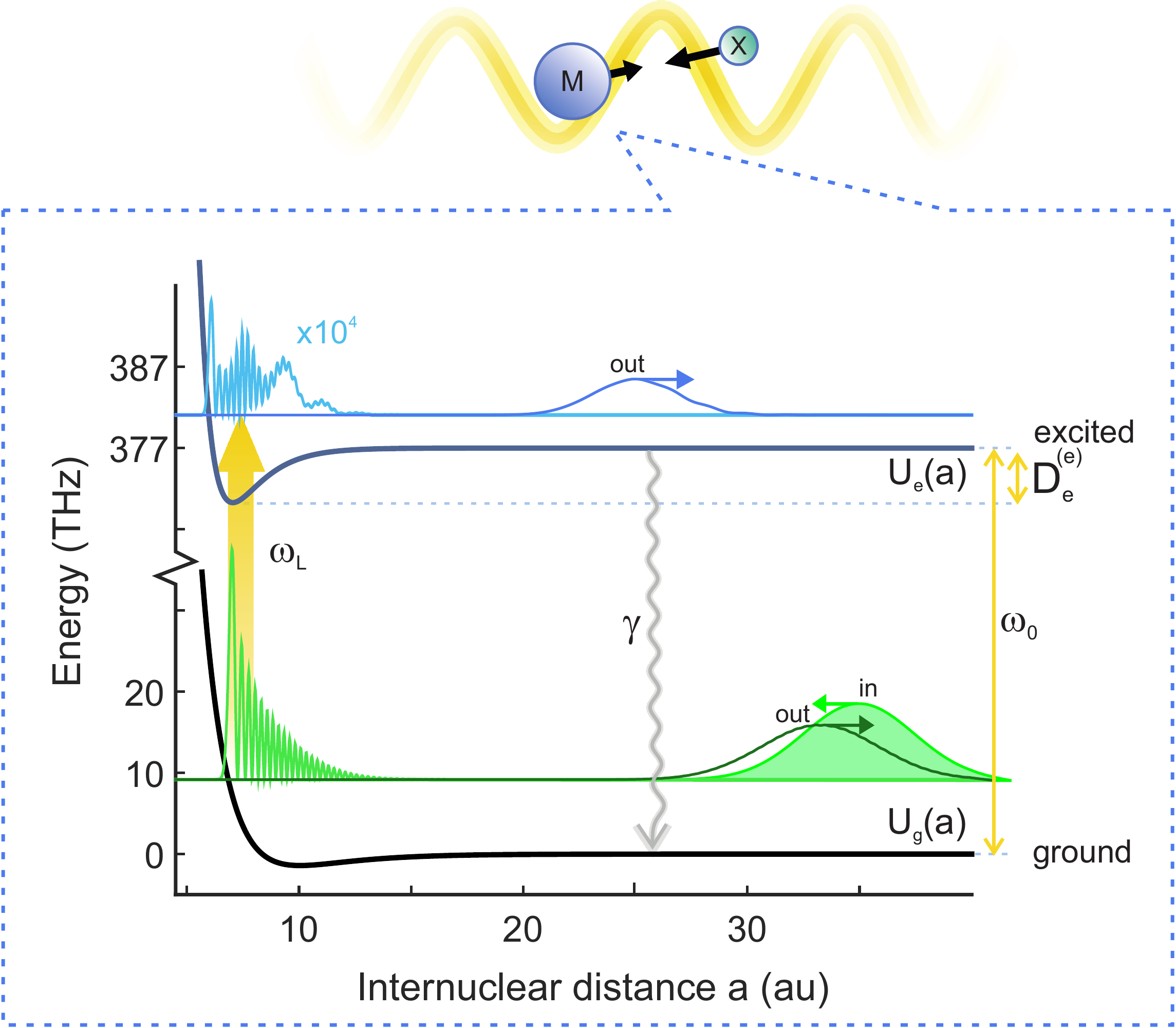}
\caption{\emph{Cooling process}. Dynamics of a M-X collision process showing a ground state X atom (modelled as a normalized Gaussian wave packet of group velocity $v_0$ and initial spread $\delta v$) approaching an atom M initially in the ground state. Around the turning point, exciplex ground-excited transitions are induced by a laser at frequency $\omega_{\text{L}}$ in a time window $\tau$. Following absorption at frequency $\omega_{\text{L}} < \omega_0$ an exciplex is formed with an excited state lifetime of $\tau_{\gamma}=\gamma^{-1}$. The outgoing wavepacket is composed of a small component containing excited state contribution owed to successful absorption of a photon (magnified in the illustration by a factor $10^4$) and a large amplitude ground state components. Spontaneous emission at rate $\gamma$ leads to an effective energy loss of $\Omega=\omega_0-\omega_{\text{L}}\leq D_e^{(e)}$.}
\label{fig2}
\end{figure}

\noindent \textbf{Model details} - We consider a hollow-core fibre~\cite{RussellJLT} of inner radius $r$ and length $\ell$ transmitting around the laser frequency $\omega_{\text{L}}$. The transmission bandwidth of the fibre is larger than $\delta \omega$ - the bandwidth of the laser. This can be realized in several types of hollow-core photonic crystal fibres (HC-PCF), where light is guided without diffraction in the empty central region by either a photonic bandgap or anti-resonance guidance such as single-ring fibre. They offer broad transmission windows and low transmission loss~\cite{Gao2018} and interaction with the dopant gas can be extremely efficient. Many experiments from nonlinear optics~\cite{RussellJLT} to quantum optics~\cite{FingerPRL} have been demonstrated in gas-filled hollow-core fibres. The fibre is filled with M atoms of mass $m_{\text{M}}$ and density $n_{\text{M}}$ and X atoms with mass $m_{\text{X}}$ and density $n_{\text{X}}$. The gas is enclosed in the volume $V=\pi r^2 \ell$ and at a given temperature $T$ it obeys the ideal gas law $p=(n_{\text{M}}+n_{\text{X}}) k_{\text{B}} T$. The laser input power $\mathcal{P}_{in}$ is assumed to be distributed constantly over the fibre radius such that the field amplitude $\mathcal{E}_{in}$ is linked to the input power by $\mathcal{E}_{in}^2=\mathcal{P}_{in}/(\epsilon_0 \pi r^2 c)$ (the group velocity is practically equal to the speed of light $c$). In Fig.~\ref{fig2} the ground and excited states of the M-X exciplex are represented with their energies $U_{g,e}$ plotted as a function of interparticle distance $a$. The illustration assumes Morse type potentials (with large depth expressed as frequency $D_e^{(e)}$ for the excited state and almost flat ground state) and roughly depicts the scattering of a Gaussian wavepacket with a relative group velocity $v$ and spread $\delta v$ representing the initial stage of the M-X collision (see Appendix for details). The laser transition is chosen for optimal Franck-Condon overlap between scattering wavefunctions off the ground and excited potentials such that $\hbar \Omega = \hbar(\omega_0-\omega_{\text{L}})\leq \hbar D_e^{(e)}$. The turning point is at $a_0$ where $U_g(a_0)=U_g(\infty)+E_{kin}$: this insures that during absorption, the speed of the exciplex is so small that the Landau-Zener tunneling parameter $e^{-\pi|\tilde{\chi}_{R}|^{2}/[2\partial_{a} U_{g}(a)v]}$ for any realistic Rabi frequencies $\tilde{\chi}_{R}$ is negligible. The cooling cycle consists of the absorption of a laser photon at $\omega_{\text{L}}$, followed by a spontaneously emitted photon at $\omega_0$ (in the long timescale $\tau_{\gamma}=\gamma^{-1}$) with an effective energy loss $\hbar \Omega$ reducing the gas kinetic energy.\\
Assuming $n_{\text{X}}\gg n_{\text{M}}$, we focus on the collision process between single M atoms and a sea of neighboring X atoms (which means we assume a small enough density of M such that M-M collisions are unlikely). The collision rate is $\kappa = n_{\text{X}} \sigma_{\text{cool}} v$, where $\sigma_{\text{cool}}$ represents a fraction of the total scattering cross section containing only events favorable to the cooling dynamics (a detailed discussion is presented in the Appendix taking into account the centrifugal barrier for off-axis collisions). We assume that the incoming particle comes at the mean thermal speed $v=\sqrt{3 k_{\text{B}} T/\mu}$ (with $\mu$ being the exciplex reduced mass). For head-on collisions, the X atom comes at a group velocity $v$ and it is roughly slowed down to $0$ during time $\tau$. This time can be  deduced from a numerical simulation of the wavepacket dynamics (see Appendix) and it corresponds to the effective laser excitation time. During $\tau\ll\tau_{\kappa}=1/\kappa$ the exciplex can resonantly absorb a laser photon and return to the bare configuration with a $\omega_0=\omega_{\text{L}}+\Omega$ transition frequency. During a much larger timescale $\tau_{\gamma} \gg \tau_{\kappa}$ a spontaneous emission process dissipates the acquired energy into free space.\\
Loss of power from the laser field occurs progressively through the fibre length. For an atom at position $z$ with a steady state excited state population $\rho_{ee}(z)$ the net power loss equals $\gamma\rho_{ee}(z)\hbar \omega_0$. The net power loss from the fibre in an infinitesimal slice of dimensions $\pi r^2 dz$ is then:
\begin{equation}
d\mathcal{P}(z)= - \gamma\hbar \omega_{\text{L}} \pi r^2 n_{\text{M}} \rho_{ee}(z)dz.
\label{dIz}
\end{equation}
\begin{figure}[t]
\includegraphics[width=0.99\columnwidth]{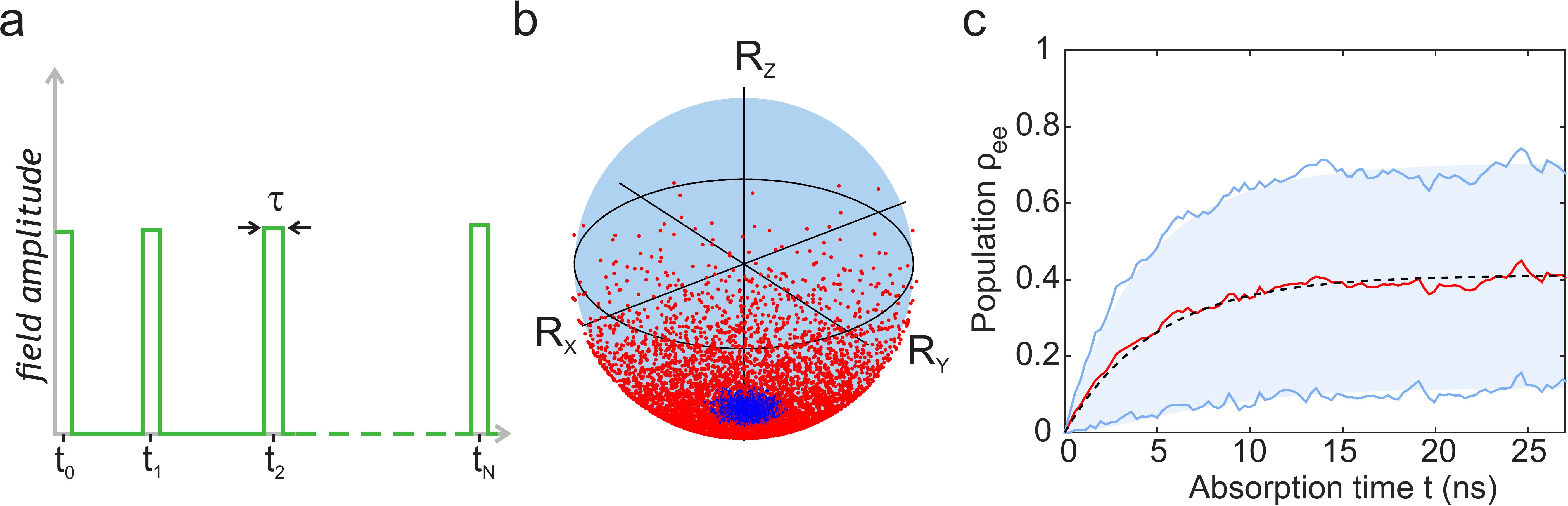}
\caption{\emph{Absorption-emission cycles}. Single M atom excitation dynamics during evolution time $\tau_{\gamma}$. a) The interaction with the laser is modeled as a sequence of $N$ randomly occurring Rabi pulses of duration $\tau$ and repetition rate $\kappa$. b) Dynamics on the Bloch sphere shows the diffusion of the initially delta-peaked spin distribution at early (blue) and later times (red). c) Comparison of average population dynamics obtained analytically (black, dashed) and numerically (red) and spin variance (blue shaded region).}
\label{fig3}
\end{figure}
Integrating the above equation with initial condition $\mathcal{P}(z=0)=\mathcal{P}_{in}$ requires knowledge of the single M atom excited state population function $\rho_{ee}(z)$ which we will model in the following.\\

\noindent \textbf{Absorption-emission cycles} - On the long timescale governed by $\tau_{\gamma}$ a single M atom located at position $z$ along the fibre is subjected to a series of randomly occurring pulses at time intervals $\{t_1,t_2,...t_N\}$ (with $N\simeq\tau_{\gamma}/\tau_{\kappa}$) each on average of duration $\tau$ (see Fig.~\ref{fig3}a). In a frame rotating at $\omega_{\text{L}}$, between pulses, the atom is freely evolving under the action of $H_0=\hbar \Omega |e\rangle \langle e|$. During a pulse, driven evolution takes place governed by the Hamiltonian $H_R=\hbar \tilde{\chi}_R (z) |g\rangle \langle e|+\hbar \tilde{\chi}^*_R (z) |e\rangle \langle g|$ where $\tilde{\chi}_R (z)= -\tilde{d}_{eg} \mathcal{E}(z) / \hbar $ is the Rabi frequency. The $\tilde{d}_{eg}$ is the reduced dipole matrix element between ground and excited states modified from the bare one by the Frank-Condon overlap. In the simulations of the Appendix we show that the typical values expected for this reduction run in the region of $20 \%$. Under typical conditions (see Appendix) $\Omega \tau_{\kappa}>1$ implies that the dynamics is well approximated by a series of $N$ phase-randomized Rabi pulses describing a random walk on the single atom Bloch sphere. A diffusion equation can be deduced for a probability distribution function $u(\varphi,\theta,t)$ of the Bloch vector (characterizing the probability of the state to point around a direction described by angles $\theta$ and $\varphi$) that reads (see Appendix):
\begin{eqnarray}
  \partial_t u &=& \frac{\mathcal{D}(z)}{\sin(\theta)}
  \left\{\partial_\theta\left[\sin(\theta)\partial_\theta \right]
    u + \frac{1}{\sin(\theta)}\partial_{\varphi\varphi}u\right\}\\\nonumber
& & -\gamma u + \frac{\gamma}{\pi}\delta(\cos(\theta)-1),
\end{eqnarray}
with a diffusion rate
$\mathcal {D} (z)= \tilde{\chi}^2_R(z) \kappa \tau^2/\pi \propto
\mathcal{P}(z)$ and the natural decay given by $\gamma$.
The solution can be written as an expansion in Legendre polynomials:
\begin{equation}
u = \sum^{\infty}_{n=0} \frac{2n+1}{4\pi}\left[\frac{(\xi_{n}(z)-\gamma)e^{-\xi_{n}(z)t}+\gamma}{\xi_{n}(z)}\right]P_{n}(\cos\theta),
\end{equation}
where $\xi_{n}(z) = \mathcal{D}(z)n(n+1)+\gamma$.
The analytical expression above has been tested against Monte-Carlo simulations of the Bloch vector dynamics $\bold{R}$ starting on the bottom of the Bloch sphere and evolving under the random action of consecutive Rabi pulses (see Fig.~\ref{fig3}b) and natural decay. Both the dynamics of the average excited state population and of its variance fit very well with the numerically predicted evolution (Fig.~\ref{fig3}c). The final population $\rho_{ee}(z)$ is obtained from the spin distribution via $\rho_{ee}(z,t)=\left(1-\int_{-1}^{1}d(\cos{\theta}) u(\theta,z,t)\cos{\theta}\right)/2$ and results in
\begin{equation}
\rho_{ee}(t,z)= \frac{\mathcal{D}(z)}{2\mathcal{D}(z)+\gamma}\left(1-e^{-(2\mathcal{D}(z)+\gamma)t}\right).
\end{equation}
The steady state value of the excited state population is given by $\rho_{ee}(z)= \mathcal{D}(z)/(2\mathcal{D}(z)+\gamma)$ which converges to $1/2$ if $\mathcal{D}(z)$ is much larger than $\gamma$.\\

\noindent \textbf{Intensity profile} - Integrating Eq.~\ref{dIz} leads to a general solution for the steady state laser power within the fibre at any point $z$. This is given by the inverse function of
\begin{equation}
z(\mathcal{P}) = \frac{(\mathcal{P}_{in}-\mathcal{P})}{\mathcal{A}} - \frac{1}{\mathcal{B}\mathcal{A}}\ln\left(\frac{\mathcal{P}}{\mathcal{P}_{in}}\right),
\label{z}
\end{equation}
with $\mathcal{P} \in \{0,\mathcal{P}_{in}\}$ where the coefficients are given by
\begin{subequations}
  \begin{align}
  \mathcal{A} &=\frac{1}{2}\hbar \omega_{\text{L}} \gamma \pi r^2 n_{\text{M}} ,\\
  \mathcal{B} &= \frac{2\tilde{d}_{eg}^{2}\sigma_{\text{cool}} \sqrt{3k_{\text{B}}}}{\hbar^2\pi^{2}\epsilon_0 \sqrt{\mu}\gamma c}\frac{\sqrt{T}n_{\text{X}} \tau^2}{r^2}.
  \end{align}
\end{subequations}
Approximate solutions for Eq.~\eqref{z} can be found in particular regimes where $\mathcal{B}\mathcal{P}_{in}\gg1$ (linear regime) or $\mathcal{B}\mathcal{P}_{in}\ll1$ (exponential regime). In the exponential regime a standard Beer-Lambert absorption law is retrieved as  $\mathcal{P}(z)=\mathcal{P}_{in} e^{-\mathcal{A}\mathcal{B}z}$. In the linear regime one strongly departs from this behavior and the power linearly decreases with distance as $\mathcal{P}(z)=\mathcal{P}_{in}-\mathcal{A}z$. In reality, for arbitrary values of $\mathcal{B}\mathcal{P}_{in}$, the power drop along the propagation direction always follows a linear law $\mathcal{P}(z)=\mathcal{P}_{in} - \mathcal{B}\mathcal{P}_{in}/(\mathcal{B}\mathcal{P}_{in}+1)\mathcal{A}z$ (for small z) and ends with an exponential fall-off $\mathcal{P}(z)=\mathcal{P}_{in}e^{\mathcal{B}(\mathcal{P}_{in}-\mathcal{A}z)}$ (for large z). In the linear regime, a penetration depth (corresponding to $\mathcal{P}({\ell_{depth}})=0$) can be defined for a long enough fibre:
\begin{eqnarray}
\ell_{depth} &=&\frac{2 \mathcal{P}_{in}}{\hbar \omega_{\text{L}} \gamma \pi r^2 n_{\text{M}}}.
\end{eqnarray}
Similarly, one can define a penetration depth in the exponential regime; however for simplicity of the analytical expressions we restrict ourselves to the linear regime. \\

\noindent \textbf{Cooling and heating processes} - The total power loss over the length of the fibre is simply expressed as
\begin{equation}
\mathcal{P}_{cool}= \frac{\Omega}{\omega_{\text{L}}}\int_{\mathcal{P}_{in}}^{\mathcal{P}(\ell)} d{\mathcal{P}}(z)=\frac{\Omega}{\omega_{\text{L}}}(\mathcal{P}_{out}-\mathcal{P}_{in}).
\end{equation}
Assuming that the total energy stored in the fibre consists only of the kinetic energy of the Rb and Ar thermal gas $E_{kin}(T)=3/2 (n_{\text{X}}+n_{\text{M}}) \pi r^2 \ell k_{\text{B}} T$ the cooling rate $\beta_{cool}(T)=T^{-1}dT/dt$ can be deduced as $\beta_{cool}(T)= -\mathcal{P}_{cool}/E_{kin}(T)$. We will focus on the linear regime where a concise expression for the cooling rate can be found
\begin{equation}
 \label{coolingrate}
\beta^{lin}_{cool}(T)\approx \frac{\hbar \Omega}{k_{\text{B}} T}\frac{n_{\text{M}}}{n_{\text{M}}+n_{\text{X}}}\gamma.
\end{equation}
Some expected scalings for the cooling rate are obtained such as: i) a linear increase with respect to the cooling efficiency parameter ${\hbar \Omega}/{(k_{\text{B}} T)}$, understood as a ratio of thermal $k_{\text{B}} T$'s extracted by a single spontaneously emitted photon, ii) the increase with photon loss rate $\gamma$ , iii) increase with higher $n_{\text{M}}$ - dopant gas pressure. However some unexpected scalings deserve analysis such as: i) a decrease with increasing buffer gas pressure and ii) independence of input laser power. The unexpected scalings stem from the initial assumption of linear power drop regime. In the opposite case of exponential decrease one can recover more intuitive scaling laws by deducing $\beta^{exp}_{cool}(T)\approx \mathcal{B}\mathcal{P}_{in} \beta^{lin}_{cool}(T)$. In such a case the input power and buffer density play an important role. However, notice that increasing these values can only improve the cooling until reaching the linear regime after which the saturation effect previously pointed out intervenes. In conclusion, for optimal cooling rates to be reached, it is desirable to ensure that, for all other parameters fixed, the product $n_{\text{X}}\mathcal{P}_{in}$ is large enough such that the linear drop regime is attained after which the loss per fiber length can be increased with an increase of $n_{\text{M}}$ only.\\
For operation under room temperature conditions we will characterize the expected heating rates stemming from the contact of the fiber walls with a $T_e=300$~K environment. Assuming a fibre of outer radius $r_e$ and fibre wall (silica) conductivity $k_g$, the expected reheating power for a length $\ell$ (see Appendix) is
\begin{equation}
\mathcal{P}_{heat}=2 \pi k_g  \frac{T_e-T}{\ln{r_e/r}}\ell.
\end{equation}
Similarly to the cooling rate in Eq.~\ref{coolingrate}, we can define a heating rate
\begin{equation}
\beta_{heat}(T)\approx  \frac{4 k_g}{3 k_{\text{B}} T (n_{\text{X}}+n_{\text{M}}) } \frac{T_e-T}{r^2 \ln{\left(r_e/r\right)}}.
\end{equation}
Assuming strong enough laser drive to ensure a linear power drop regime, we can estimate a maximum reachable temperature drop at the inner fibre wall from a power balance $\mathcal{P}_{cool}+\mathcal{P}_{heat}=0$ and under the assumption that the fibre length is matched on the penetration depth $\ell_{depth}$:
\begin{equation}
(\delta T)_{max}= \gamma n_{\text{M}} \frac{\hbar \Omega }{4 k_g}r^2 \ln{\left(\frac{r_e}{r}\right)}.
\end{equation}

\begin{figure}[t]
\includegraphics[width=0.94\columnwidth]{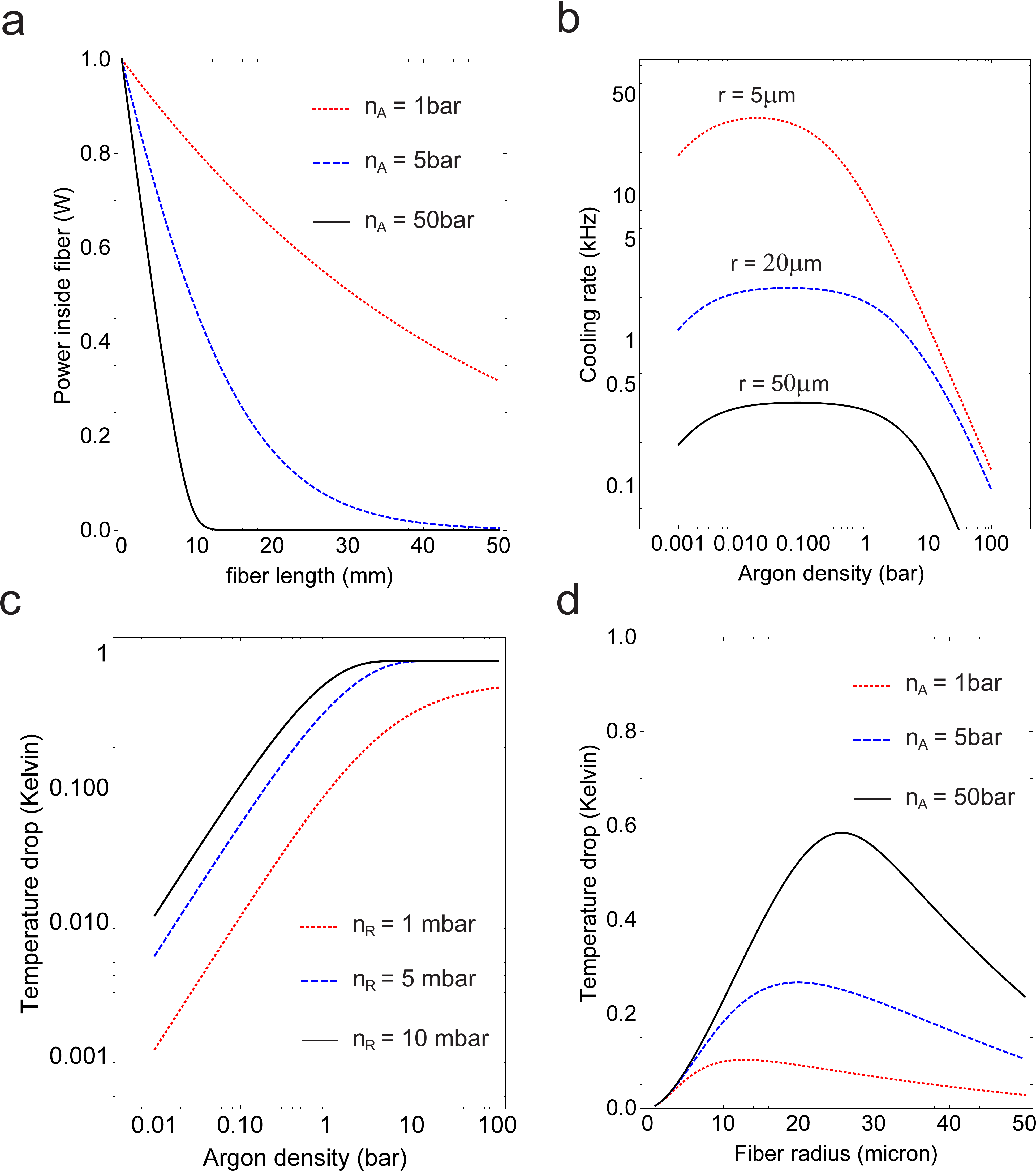}
\caption{\emph{Power, cooling rate and temperature drop}. a) Power profile along the fibre showing the transition from an exponential decay (red-dotted line) to a linear drop (black line) for increasing Ar pressure. b) The cooling rate shows an optimal region with respect to the Ar density for a fixed Rb density. This indicates that higher collision rates do not necessarily result in more efficient cooling as the mass of the gas to be cooled increases as well. c, d) The temperature drop shows a monotonous increase with Rb density and an optimum for increasing Ar density.}
\label{fig4}
\end{figure}

\begin{figure}[t]
\includegraphics[width=0.98\columnwidth]{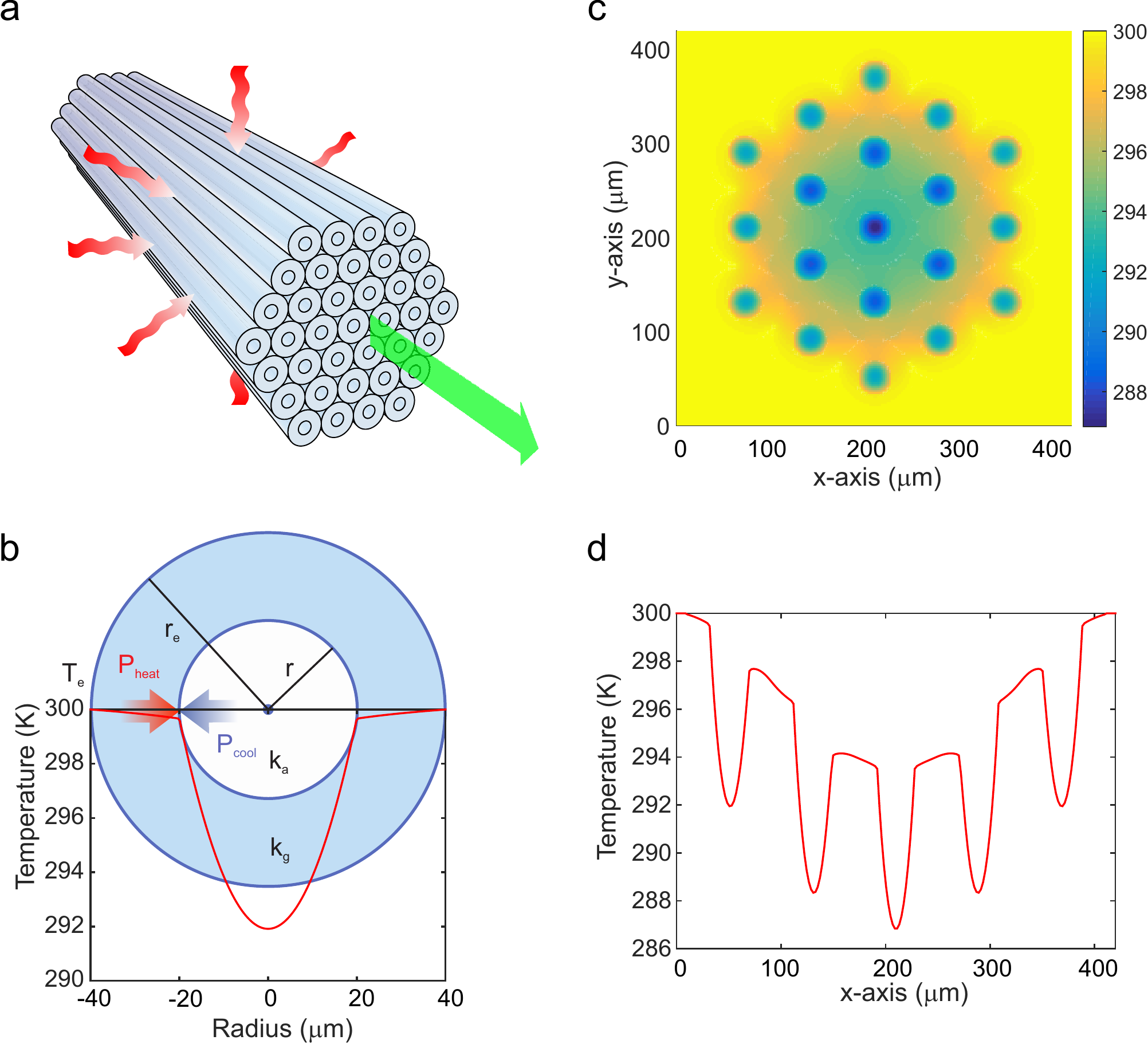}
\caption{\emph{Fibre bundle}. a) Sketch of a bundle of $N$ fibres as an arrangement that minimizes heat flow while maximizing cooling power leading to a temperature drop scaling very favorably with $N$. b) Temperature distribution for a single fibre. c) Density plot temperature map for a bundle of 19 fibres. d) Radial plot of temperature distribution within the core and cladding for the 19 fibre arrangement.}
\label{fig5}
\end{figure}

\noindent \textbf{Numerical illustration} - We focus our numerical analysis to the Rb-Ar exciplex~\cite{Baylis1969semiempirical,Pascale1973excited,Dhiflaoui2012electronic}, for which experimental results indicating cooling have been shown in Refs.~\cite{Weitz2009laser,Sass2011laser}. We list the fixed parameters in the text while tabulating derived quantities in Table~\ref{tab1}. The reduced mass for Rb-Ar collisions is $\mu=4.5112\times10^{-26}\,$kg and the corresponding cross section is $\sigma=572\,$\AA$^2$~\cite{Rosin1935effective} (a reduced effective $\sigma_{\text{cool}}=20\,$\AA$^2$ is used in the simulations - see Appendix for details). For an initial temperature $T=300$~K the resulting relative collision velocity is around $v=524\,$m/s. Here, we focus on the $\mathrm{X}^{2}\Sigma \leftrightarrow \mathrm{A}^{2}\Pi_{1/2}$ transition which in the asymptotic limit $r\rightarrow \infty$ converges to the Rubidium D1 line. The energy loss takes place at $\gamma=2 \pi \times 5.75$~MHz and the loss per cycle is approximately equal to the depth of the Rb-Ar excited state $\Omega=2 \pi \times 6.7154\,$THz. The bare ground-excited frequency difference is $\omega_0=2 \pi \times 377\,$THz and the dipole matrix element is $d_{eg}=2.537\times 10^{-29}$~Cm. Wavepacket dynamics simulated in the Appendix for collisions followed by scattering back into unbound states prove the cooling effect as a reduction of the outgoing wavepacket's velocity; they also predict a reduction in the effective dipole matrix element. This is owed to a reduction of the Frank-Condon overlap between unbound states and the incoming wavepacket. However, bound states (with larger Franck-Condon overlaps) participate as well and can be phenomenologically included by assuming the full value $d_{eg}$ in the excitation process. The mechanism leading to the escape of particles from bound states is associated with the Ar-Ar collisions in the gas leading to a quick thermalization of the bound states. A theoretical approach to be considered in the future will see the extension of numerical simulations based on coherent wavepacket dynamics (as shown in the Appendix) to a stochastic model including Brownian noise source terms governing the dynamics of the incoming Ar particle. The equilibrium distance for the excited state potential is $a_0=3.731\,$\AA. For an ideal gas at room temperature, a pressure of $1$ bar consists of $n_0=2.414\times10^{19}/$cm$^3$. This allows us to express the considered gas densities as $n_{\text{R}}$ of the order of $1\,$mbar (realized in Ref~\cite{Weitz2009laser}) and $n_{\text{A}}$ ranging from $1\,$bar to $100\,$bar. Regarding fibre geometry, we consider fibres of $r=20\mu$m and $r_e=70\mu$m with a listed thermal conductivity (silica) $k_g\approx0.8$~WK$^{-1}$m$^{-1}$. The thermal conductivity for Argon is given by $k_a\approx0.03$~WK$^{-1}$m$^{-1}$. For increasing Ar density the cross-over between the exponential and the linear regime is illustrated in Fig.~\ref{fig4}a. The dependence on Ar density is further explored in Fig.~\ref{fig4}b, where the cooling rate shows a plateau-like behavior for moderate pressures for all fibre radii considered. The saturation of the cooling rate with increasing noble gas pressure is owed to the increase of the gas mass to be cooled; as the penetration depth is reached, the volume $\ell-\ell_{depth}$ contains gas that does not participate in the cooling. Figures.~\ref{fig4}c,d explore the behavior of the achievable temperature drop at the cladding-core interface with varying densities and fibre radius. For a given Rb density an optimal Ar density is always reached: at moderate Rb densities around $1$~mbar the optimal Ar density surpasses $100$~bar, however for not larger than $10$~mbar Rb density, a much smaller Ar density around $10$~bar suffices. For varying fibre radia an optimal Ar density is also found: as the volume of the gas to be cooled is increased higher densities are required.\\
\begin{table}[b]
\begin{tabular}{ l| l | l | l }
\hline
Symbol & Description & Value  & Units\\ \hline
$\kappa$ & collision rate &  $5.83$ & GHz \\ \hline
$\tau$ & absorption time ($T=300$K)& $1$ & ps \\ \hline
$\tau_{\kappa}$ & average collision time  & $171.61$ & ps \\ \hline
$\tau_{\gamma}$ & spontaneous emission time & $27.68$ & ns \\ \hline
$\sigma_{\text{cool}}$ & cooling cross-section & $20$ & $\AA^2$ \\ \hline
$\mathcal{D}(z=0)$ & maximum diffusion rate &  $33.65$ & MHz \\ \hline
\end{tabular}
\caption{\emph{Fibre cooling parameters for Rb-Ar exciplex at buffer pressure $n_{\text{X}}=10\,$bar.}}
\label{tab1}
\end{table}
While the predicted temperature drop at the cladding-core interface is of the order of Kelvins under realistic conditions, one can minimize the effect of environmental heating by designing geometries (as depicted in Fig.~\ref{fig5}a) that optimize the volume over surface ratio. Let us consider a scenario where a spatial light modulator is used to couple light into a multitude of cores of a fibre bundle containing $N$ fibres. We then simulate the heat equation with sources (the cooling power from inside the fibre cores) imposing that the outer boundary of the fibre bundle is kept at the ambient temperature of $300$~K. We plot the results of the simulation in Fig.~\ref{fig5}c as a temperature map across the fibre bundle and as a radial temperature drop in Fig.~\ref{fig5}d. Comparison with the single fibre case of Fig.~\ref{fig5}b illustrates an improved temperature drop in the inner core. More generally, for large $N$ we expect an improvement of the average temperature drop of the inner fibre to roughly scale inbetween $\propto\sqrt{N}$ and $N$.\\

\noindent \textbf{Conclusions/Discussions.} - We have developed a general theory for collision-assisted cooling of fibre-embedded dopant-buffer gas mixtures of exciplexes. We have identified the role of dopant/buffer gas densities, geometry of the fibre, input laser power and specifics of the exciplex potential landscape. Our results predict observable temperature drops under ambient conditions in single fibre and especially in fibre-bundle geometries. The model considered here disregards a few effects such as: off-resonant laser light absorption, inelastic collisions, intrinsic fibre losses, three-body collisions, spontaneously emitted photon reabsorption within the gas, superfluorescence. Some of these effects are purely detrimental but controllable. For example, off-resonant absorption is small as the cross-section scales as $\propto (\tilde{\chi}_R(z)/\Omega)^2\ll1$. Intrinsic fibre losses are leading to not more than a factor of $10^{-6}$ in power loss for our cm-long assumed fibres (negligible compared to the collision-induced loss). Inelastic collisions, leading to the relaxation of the exciplex without loss of kinetic energy have already been estimated as significantly smaller than elastic ones thus negligible~\cite{Speller1979quenching}. Other effects such as three-body collisions could be beneficial as they could provide additional mechanisms for cooling: for deep excited state potentials, where bound exciplexes are likely, scattering off additional X atoms can destroy the bound state leading to the release of X atoms and consequent cooling. This effect will be theoretically considered in the near future in the framework of a stochastic model describing the diffusive Brownian motion dynamics of the buffer gas particle. The reabsorption of spontaneously emitted photons within the gas along the fibre could also provide a beneficial mechanism for cooling as it would lead to superradiant/superfluorescence and practically increase the cooling rate factor of $\gamma$ to an improved $\gamma_{sf}\gg\gamma$ (for the high optical depth assumed here). The drawback is an increased photon recoil momentum which however sets a much lower bound than the expected temperature limit aimed in our analysis.
A more serious limitation arises from the analysis of the potential energy landscape in the excited state.\\

\noindent \textbf{Acknowledgements} - We acknowledge very useful discussions with Martin Weitz and Guido Pupillo. This work is supported by the Max Planck Society (C. S and C.G.).

\bibliography{collisionsbib}
\clearpage
\newpage
\section*{Appendices}

\subsection{Wave-packet dynamics}
A realistic model of the absorption-emission cycle during collisions between an alkali atom and a rare-gas atom needs the solution of the Schr\"odinger equation
\begin{eqnarray}
\label{Ap:Coll:Eq1}
i\hbar\frac{\partial}{\partial t} \ket{\psi} &=& H \ket{\psi}
\end{eqnarray}
that includes the dynamics of the system. This is included by considering $\ket{\psi} = \psi_{g}(\mathbf{r},t)\ket{g} + \psi_{e}(\mathbf{r},t)\ket{e}$, where $\psi_{j}(\mathbf{r},t)$ with $j\in \{g,e\}$ are the ground and excited state wavefunctions, respectively, which depend on the relative distance between the alkali and the rare-gas atom.
The Hamiltonian is given by
\begin{eqnarray}
\label{Ap:Coll:Eq2}
H &=& -\frac{\hbar^2}{2\mu}\Delta + \hbar U_{g}(r)\ket{g}\bra{g} + \hbar U_{e}(r)\ket{e}\bra{e}\\\nonumber
& & - \mathbf{d}_{eg}\mathbf{E}(t)\ket{e}\bra{g} - \mathbf{d}^{*}_{eg}\mathbf{E}(t)\ket{g}\bra{e},
\end{eqnarray}
where $\mathbf{E}(t) = (1/2)\mathbf{E}_{0}e^{i\omega_{L}t} + c.c.$ describes the laser field (under dipolar approximation assuming that the dynamics takes place within a wavelength). Here, we ignore Doppler shifts since these are by three orders of magnitude smaller than the detunings that result from the molecular potentials. The detuning of the laser frequency with respect to the bare atomic transition chosen in numerical simulations is $\omega_{0}-\omega_{L} = 2\pi \times 5.2\,$THz, which is close to the depth of the excited potential with $D^{(e)}_{e} = 2\pi \times 6.72\,$THz. Realistic models for the excited and ground potentials are given by Morse potentials
\begin{equation}
U_{j}(r) = D^{(j)}_{e}\left[1-e^{-a_{j}(r-r_{e,j})}\right]^{2}
\end{equation}
where $D^{(j)}_{e}$ is the depth and $a_{j}$ is the width of the potential with $j \in \{g,e\}$. The parameters for a Rubidium-Argon collision are taken from Ref.~\cite{Dhiflaoui2012electronic} and are exactly illustrated in Fig.~\ref{fig2} in the main text. For $r \rightarrow \infty$ the potentials converge to the bare energies of the Rb atom.
\begin{figure}[t]
\includegraphics[width=0.98\columnwidth]{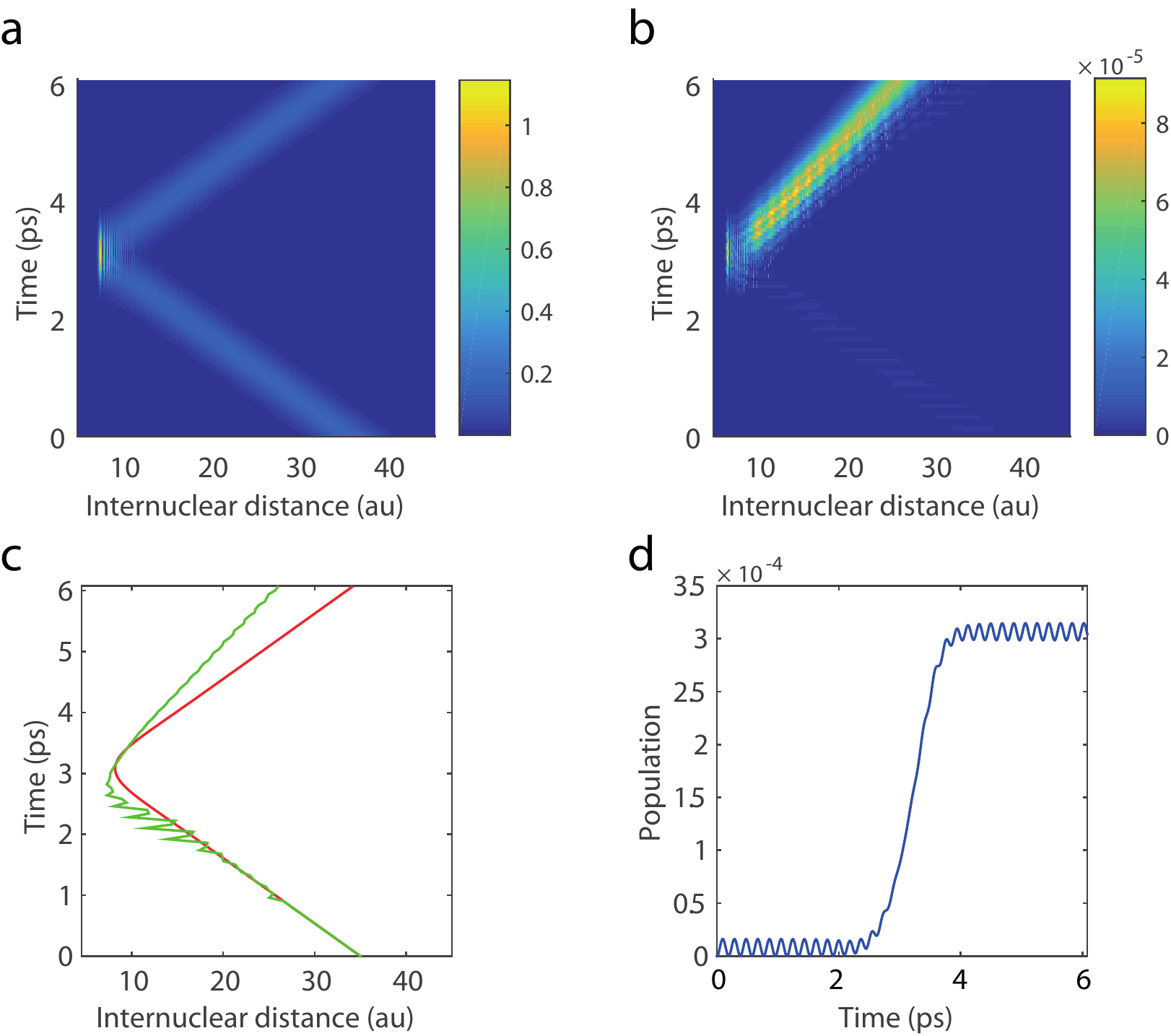}
\caption{\emph{Collision induced excitation process}. a) Evolution of a wave packet in the ground state. b) Evolution of the excited state. Initial small ripples correspond to off-resonant weak excitation in the incoming part. Large (normalized by $10^{-5}$ excitation can be observed after the interaction with the laser field. Notice that the slope of the outgoing excited state is smaller than of the outgoing ground state indicating loss of kinetic energy following collision-aided absorption of a laser photon. c) The expectation value for the distance between the particles is plotted for the ground (red line) and excited state (green line) showing a reduction in the kinetic energy of the outgoing excited state wave packet (smaller slope). d) Simulation of the excited state population dynamics before during the whole collision process. Small ripples show the small off-resonant Rabi oscillations while the resonant passing through the laser excitation region shows a large population accumulation in the outgoing wavepacket.}
\label{fig2A}
\end{figure}
Using the transformation $\psi_{g}(\mathbf{r},t) = \phi_{g}(\mathbf{r},t)$, $\psi_{e}(\mathbf{r},t) = e^{-i\omega_{L}t}\phi_{e}(\mathbf{r},t)$ to the rotating frame of the light field and by omitting the fast rotating components (rotating wave approximation) we derive a set of coupled equations
\begin{subequations}
  \begin{align}
	\label{Ap:Coll:Eq3}
	i\hbar\dot{\phi}_{g}(\mathbf{r},t) = -\frac{\hbar^2}{2\mu}\Delta\phi_{g}(\mathbf{r},t) + \hbar U_{g}(r)\phi_{g}	(\mathbf{r},t)
	+ \frac{\hbar\chi_{R}}{2}\phi_{e}(\mathbf{r},t)\\\nonumber
	i\hbar\dot{\phi}_{e}(\mathbf{r},t) = -\frac{\hbar^2}{2\mu}\Delta\phi_{e}(\mathbf{r},t) + (\hbar U_{e}(r)-\hbar\omega_{L})	\phi_{e}(\mathbf{r},t) \\
	+ \frac{\hbar\chi^{*}_{R}}{2}\phi_{g}(\mathbf{r},t),
	\end{align}
\end{subequations}
with $\chi_{R} = -\mathbf{d}^{*}_{eg}\mathbf{E}_{0}/\hbar$.
For head on collision (neglecting the total angular momentum of the two atom system) we obtain the equations
\begin{subequations}
  \begin{align}
\label{Ap:Coll:Eq4}
i\hbar\dot{\xi}_{g}(r,t) = -\frac{\hbar^2}{2\mu}\frac{\partial^2}{\partial r^2}\xi_{g}(r,t) + \hbar U_{g}(r)\xi_{g}(r,t)
+ \frac{\hbar\chi_{R}}{2}\xi_{e}(r,t)\\\nonumber
i\hbar\dot{\xi}_{e}(r,t) = -\frac{\hbar^2}{2\mu}\frac{\partial^2}{\partial r^2}\xi_{e}(\mathbf{r},t) + (\hbar U_{e}(r)-\hbar\omega_{L})\xi_{e}(r,t)\\
+ \frac{\hbar\chi^{*}_{R}}{2}\xi_{g}(r,t),
\end{align}
\end{subequations}
which we solve numerically. These expressions can be obtained by separating the components of the wavefunctions $\phi_{j}(\mathbf{r},t) = (\xi_{j}(r,t)/r)Y_{0,0}(\theta,\varphi)$ with $j\in \{g,e\}$. The initial wave packet is described by an incoming Gaussian wavepacket
\begin{equation}
\xi_{g}(r,0) =  \sqrt{\frac{\Gamma}{\sqrt{\pi}}}e^{-\frac{\Gamma^2(r-r_0)^2}{2}}e^{-i(k_{0}(r-r_0))}
\end{equation}
where $\int^{\infty}_{0}|\xi_{g}(r,0)|^{2}dr = 1$.  The wave packet used in the simulation that is presented in Fig.~\ref{fig2A} has an average kinetic energy $E_{av} = \hbar^2 k^{2}_{0}/(2\mu) = (3/2)k_{B}T$ at $T=300\,$K and a narrow spread in momentum space.

\subsection{Absorption-emission cycles}
On the long timescale governed by $\tau_{\gamma}$ a single Rb atom located at position $z$ along the fibre is subjected to a series of randomly occurring pulses at time intervals $\{t_1,t_2,...t_N\}$ (with $N\simeq\tau_{\gamma}/\tau_{\kappa}$) each on average of duration $\tau$ (see Fig.~\ref{fig3}a). In a frame rotating at $\omega_{\text{L}}$, between pulses, the atom is freely evolving under the action of $H_0=\hbar \Omega |e\rangle \langle e|$. During a pulse, driven evolution takes place governed by the Hamiltonian $H_R=\hbar \tilde{\chi}_R (z) |g\rangle \langle e|+\hbar \tilde{\chi}^*_R (z) |e\rangle \langle g|$ where $\tilde{\chi}_R (z)= \tilde{d}_{eg} \mathcal{E}(z) / \hbar $ is the Rabi frequency. Under typical conditions (see Appendix) $\Omega \tau_{\kappa}>1$ implying that the dynamics is well approximated by a series of $N$ on average equal area and phase-randomized Rabi pulses describing a random walk on the single atom Bloch sphere. A diffusion equation $\partial_t u = \mathcal{D}\Delta u - \gamma u + f$ can be deduced for a probability distribution function $u(\varphi,\theta,t)$ of the Bloch vector (characterizing the probability of the state to point around a direction described by angles $\theta$ and $\varphi$) that reads:
\begin{eqnarray}
\label{Ap:Eq.1}
\partial_t u &=& \frac{\mathcal{D}(z)}{\sin(\theta)}
\left\{\partial_\theta\left[\sin(\theta)\partial_\theta
  \right]
  u + \frac{1}{\sin(\theta)}\partial_{\varphi\varphi}u\right\}\\\nonumber
& & -\gamma u + \frac{\gamma}{\pi}\delta(\cos(\theta) -1),
\end{eqnarray}
with a diffusion rate
\begin{equation}
\label{Ap:Eq.2}
\mathcal {D} (z)= \frac{1}{\pi}\tilde{\chi}^2_R \kappa \tau^2 = \frac{1}{\pi} \left(\frac{d_{eg} \tau}{\hbar} \right)^2 \frac{\kappa \mathcal{P}(z)}{\epsilon_0 \pi r^2 c}.
\end{equation}
To solve this equation we perform a separation of variables $u(\varphi, \theta, z, t) = \zeta(\varphi)P(\theta)\eta(t,z)$ and consider the homogeneous differential equation $\partial_t u = \mathcal{D}\Delta u - \gamma u$ at first.
With the initial population being in the ground state $\ket{g}$, which we identify with the coordinate $\theta = 0$ (the excited state $\ket{e}$ is identified with $\theta = \pi$) and assuming that the diffusion progresses equaly in all directions we can choose the solution that is independent of the azimuthal angle given by $\zeta(\varphi) = 1/2\pi$.\\
The remaining differential equations are given by
\begin{subequations}
  \begin{align}
	\label{Ap:Eq.3}
	\partial_t \eta(t,z) = -\lambda(z)\eta(t,z) \; \\
	\label{Ap:Eq.4}
	\partial_x((1-x^2)\partial_x)P(x) = -\frac{(\lambda(z)-\gamma)}{\mathcal{D}(z)}P(x),
	\end{align}
\end{subequations}
where we have used the differential relation $-d(\cos(\theta)) = \sin(\theta)d\theta$ and defined $x = \cos(\theta)$.
The finite solutions for Eq.~\ref{Ap:Eq.4} are Legendre polynomials which set $(\lambda(z)-\gamma)/\mathcal{D}(z) = n(n+1)$ with $n \in \mathbb{N}_{0}$ and we define $\xi_{n} = \mathcal{D}(z)n(n+1)+\gamma$. This allows us to postulate an ansatz for the general solution given by
\begin{eqnarray}
\label{Ap:Eq.4a}
u(x,z,t) = \frac{1}{2\pi}\sum_{n=0}^{\infty} w_{n}(t,z)P_{n}(x).
\end{eqnarray}
With $\delta(x-1) = \sum_{n=0}^{\infty}A_{n}P_{n}(x)$ where
\begin{eqnarray}
\label{Ap:Eq.6}
A_{n} &=& \frac{\int_{-1}^{1}\delta(x-1)P_{n}(x)dx}{\int_{-1}^{1} P_{n}(x)^2 dx} = \frac{2n+1}{4}.
\end{eqnarray}
For the nonhomogeneous differential equation Eq.~\ref{Ap:Eq.1} we obtain
\begin{eqnarray}
\label{Ap:Eq.6a}
\nonumber
\gamma\sum_{n=0}^{\infty}\left(\frac{2n+1}{2}\right)P_{n} &=& \sum_{n=0}^{\infty}[\partial_{t}w_{n} + \xi_{n}w_{n}]P_{n},\\
\end{eqnarray}
which results in differential equations $\dot{w_{n}}(t,z) + \xi_{n}w_{n}(t,z) = ((2n+1)/2)\gamma$ for each $n \in \mathbb{N}_{0}$. With $u(x,z,0) = \frac{1}{\pi}\delta(x-1)$ (since $\int_{0}^{2\pi}\int_{-1}^{1}u(x,z,0)dx d\varphi = 1$) we obtain
\begin{equation}
\label{Ap:Eq.6b}
w_{n}(t,z) = \left(\frac{2n+1}{2} \right)\left[\frac{\gamma}{\xi_{n}(z)}\left(1-e^{-\xi_{n}(z)t}\right) + e^{-\xi_{n}(z)t} \right]
\end{equation}
Since the Bloch sphere harbors the relation $\langle \cos(\theta) \rangle = \rho_{gg} - \rho_{ee}$ and
\begin{eqnarray}
\label{Ap:Eq.8}
\langle \cos(\theta) \rangle &=& \int_{0}^{2\pi}\int_{0}^{\pi} \cos(\theta)u(\theta,z,t)\sin(\theta)d\theta d\varphi \\\nonumber
&=& \frac{2}{3}w_{1}(t,z),
\end{eqnarray}
we can derive for the average population in the excited state
\begin{eqnarray}
\label{Ap:Eq.9}
\rho_{ee}(z,t) &=& \frac{\mathcal{D}(z)}{2\mathcal{D}(z)+\gamma}\left(1-e^{-(2\mathcal{D}(z)+\gamma)t}\right).
\end{eqnarray}
This is in good agreement with numerical Monte Carlo simulations (see Fig.\ref{fig3}b,c) where we find that the final state population $\rho_{ee}(z)$ at $t\rightarrow \infty$ is given by
\begin{equation}
\label{Ap:Eq.10}
\rho_{ee}(z)= \frac{\mathcal{D}(z)}{2\mathcal{D}(z)+\gamma}.
\end{equation}
The standard deviation of the population can be obtained from calculating the variance
\begin{eqnarray}
\label{Ap:Eq.11}
\nonumber
\langle \cos(\theta)^2 \rangle - \langle \cos(\theta) \rangle^2 &=& \frac{2}{3}\left[\frac{6\mathcal{D}e^{-(6\mathcal{D}+\gamma)t}+\gamma}{(6\mathcal{D}+\gamma)}\right] + \frac{1}{3}\\\nonumber
& & -\left[\frac{\left(\gamma + 2\mathcal{D}e^{-(2\mathcal{D}+\gamma)t}\right)^2}{(2\mathcal{D}+\gamma)^{2}}\right]
\end{eqnarray}
This results in the standard deviation
\begin{eqnarray}
\label{Ap:Eq.12}
\Delta\rho_{ee}(z,t) = \frac{1}{2}\sqrt{\langle \cos(\theta)^2 \rangle - \langle \cos(\theta) \rangle^2}
\end{eqnarray}

\subsection{Derivation of the diffusion equation}
The justification that a diffusion equation gives a good representation of the dynamics is given in this subsection.
The dynamical evolution on the Bloch sphere in the resonant case with a Rabi frequency $\chi_{R}$ is described by the differential equation $\dot{\pmb{R}} = \pmb{\chi}_{R}\times \pmb{R}$.
For a parametrization of the Blochsphere given by $(\theta, \varphi) \in [0,\pi)\times[0,2\pi)$ we obtain the relation
\begin{equation}
\label{Ap:Diff.Eq.2}
d\theta = -\tilde{\chi}_{R}\sin(\varphi)dt,
\end{equation}
for the change of the altitude angle $\theta$ and its dependence on the azimuthal angle $\varphi$.
In the case of randomly occurring collisions inducing resonant Rabi-pulses of average duration $\tau$ with a repetition (collision) rate $\kappa$, where $\tau \ll \tau_{\kappa} = 1/\kappa$ we can have a short excitation or de-excitation by $\Delta \theta = \tilde{\chi}_{R} \tau$. This is followed by a longer average phase-randomization time $\tau_{\kappa}-\tau$ due to the evolution at a large detuning given by $\hbar\Omega$ after the collision event. Due to the phase randomization we find the state with equal probability at any angle $\varphi$ after an average collision time $\tau_{\kappa}$. Therefore, for every pump-time step $\tau$ and by using Eq.~\ref{Ap:Diff.Eq.2} we find the probability to go up or down along the altitude angle $\theta$ to be $p_{\pm} = 1/\pi\pm f(\theta)$, respectively, while the probability for remaining at the same altitude angle is given by $p_{0} = 1-2/\pi$.\\ The term $f(\theta) = \frac{\tilde{\chi}_{R}\tau}{2\pi}\cot(\theta)$ is a geometric correction which vanishes at the equator of the Bloch sphere and emerges due to the curvature of the sphere. Here, $\cot(\theta)$ is the value of the Christoffel symbol $\Gamma^{\varphi}_{\theta \varphi}$ of the Levi-Civita connection of the Riemanian metric of the sphere and gives the change of the covariant bases vector along the altitude $\pmb{e}_{\theta}$ with the azimuthal angle $\varphi$ projected along the azimuthal bases vector $\pmb{e}_{\varphi}$ and thereby decribes the size changes of the azimuthal rings for different altitude angles.\\
This allows us to write a probabilistic equation for the state location with respect to the altitude after a collision process of time $\tau_{\kappa}$
\begin{eqnarray}
\label{Ap:Diff.Eq.3}
\nonumber
\Phi_{n+1}(m) &=& p_{+}\Phi_{n}(m+1) + p_{0}\Phi_{n}(m) + p_{-}\Phi_{n}(m-1)\\\nonumber
&=& \left(\frac{1}{\pi}+\frac{\Delta\theta}{2\pi}\cot(m\Delta\theta) \right)\Phi_{n}(m+1)\\\nonumber
& & + \left(1-\frac{2}{\pi}\right)\Phi_{n}(m)\\
& & + \left(\frac{1}{\pi}-\frac{\Delta\theta}{2\pi}\cot(m\Delta\theta) \right)\Phi_{n}(m-1).
\end{eqnarray}
Here, $n$ is the counter for the collision events and is equivalent to the time coordinate. the parameter $m$ gives the altitude location which changes stepwise by $\Delta\theta$. The difference of the state-location distribution $\Phi$ with respect to time is given by
\begin{eqnarray}
\label{Ap:Diff.Eq.4}
\nonumber
\Phi_{n+1}(m) - \Phi_{n}(m) &=& \frac{1}{\pi}\Big(\Phi_{n}(m+1)+\Phi_{n}(m-1)\\\nonumber
& & -2\Phi_{n}(m) +\frac{\Delta\theta}{2}\cot(m\Delta\theta)  \\\nonumber
& &  \times \left(\Phi_{n}(m+1)-\Phi_{n}(m-1)\right)\Big) \\
\tau_{\kappa}\partial_{t}\Phi &\cong& \frac{\Delta \theta^2}{\pi}\left(\partial_{\theta\theta}\Phi  + \cot(\theta)\partial_{\theta}\Phi \right).
\end{eqnarray}
This results in
\begin{eqnarray}
\label{Ap:Diff.Eq.5}
\partial_{t}\Phi =\left(\frac{\tilde{\chi}_{R}^2\kappa\tau^2}{\pi}\right)\frac{1}{\sin(\theta)}\partial_{\theta}(\sin(\theta)\partial_{\theta}\Phi).
\end{eqnarray}
Here, the diffusion constant is defined by $\mathcal{D}= \left(\frac{\tilde{\chi}_{R}^2\kappa\tau^2}{\pi}\right)$.
By considering a constant distribution for the azimuthal angle $\zeta(\varphi) = 1/2\pi$ due to phase randomization we obtain the full diffusion equation for the distribution $u(\varphi,\theta,t) = \zeta(\varphi)\Phi(\theta,t)$
\begin{eqnarray}
\label{Ap:Diff.Eq.6}
\partial_{t}u = \mathcal{D}\left\{\frac{1}{\sin(\theta)}\partial_{\theta}(\sin(\theta)\partial_{\theta}u)+\frac{1}{\sin^2(\theta)}\partial_{\varphi\varphi}u \right\}.
\end{eqnarray}
Using $p_{0} = 1-2/\pi-p_{\gamma}$ instead of $1-2/\pi$, where $p_{\gamma} = \gamma\tau_{\kappa}$ and by considering the accumulation of population via decay from every altitude location different than $\theta = 0$, at the groundstate $\Phi_{n}(0)$ described by the additional factor $\gamma\tau_{\kappa}\delta_{0m}$, we can find by analogous steps as has been used in this paragraph, the nonhomogeneous diffusion equation
\begin{eqnarray}
\label{Ap:Diff.Eq.7}
\partial_{t}u &=& \mathcal{D}\left\{\frac{1}{\sin(\theta)}\partial_{\theta}(\sin(\theta)\partial_{\theta}u)+\frac{1}{\sin^2(\theta)}\partial_{\varphi\varphi}u \right\}\\\nonumber
& & -\gamma u + \frac{1}{\pi}\delta(\cos(\theta)-1)
\end{eqnarray}
that incorporates radiative decay to the ground state.

\subsection{Temperature distribution for hollow core fibres - environmentally-induced heating rate}

Let us assume that laser cooling results in an instantaneous temperature $T$ constant over the inner wall of the glass encompassing the gas inside the fibre. The outer shell (with heat conductivity constant $k_g$ for silica) at $r_e$ is kept at ambient temperature $T_e$. Writing a heat equation in cylindrical coordinates for the variable $\rho$ such that $r \leq \rho \leq r_e$ in steady state and assuming independence of the temperature on angle and coordinate along the fibre axis ($\mathcal{B}\mathcal{P}_{in} \ll 1$):
\begin{equation}
\partial_{\rho} \left[\rho\partial_\rho T(r)\right] = 0,
\end{equation}
we obtain a straightforward solution
\begin{eqnarray}
\label{Ap:Eq.14}
T(\rho) &=& -C\ln\left(\frac{r_e}{\rho}\right) + T_e.
\end{eqnarray}
Since we observe the heat flux through the cylindrical area with radius $r$ given by $\mathcal{P}_{heat} = 2\pi \rho \ell k_g\partial_\rho T |_{\rho = r}$ (von Neumann boundary condition) we obtain the relation
\begin{eqnarray}
\label{Ap:Eq.15}
\rho\partial_\rho T|_{\rho=r} &=& C = \frac{P_{heat}}{2\pi l},
\end{eqnarray}
which results in
\begin{equation}
\mathcal{P}_{heat} = 2 \pi k \ell \frac{T_e-T(r)}{\ln{r_e/r}},
\end{equation}
describing the incoming heat which effects the temperature of the inner gas mixture.
\begin{figure}[t]
\includegraphics[width=0.60\columnwidth]{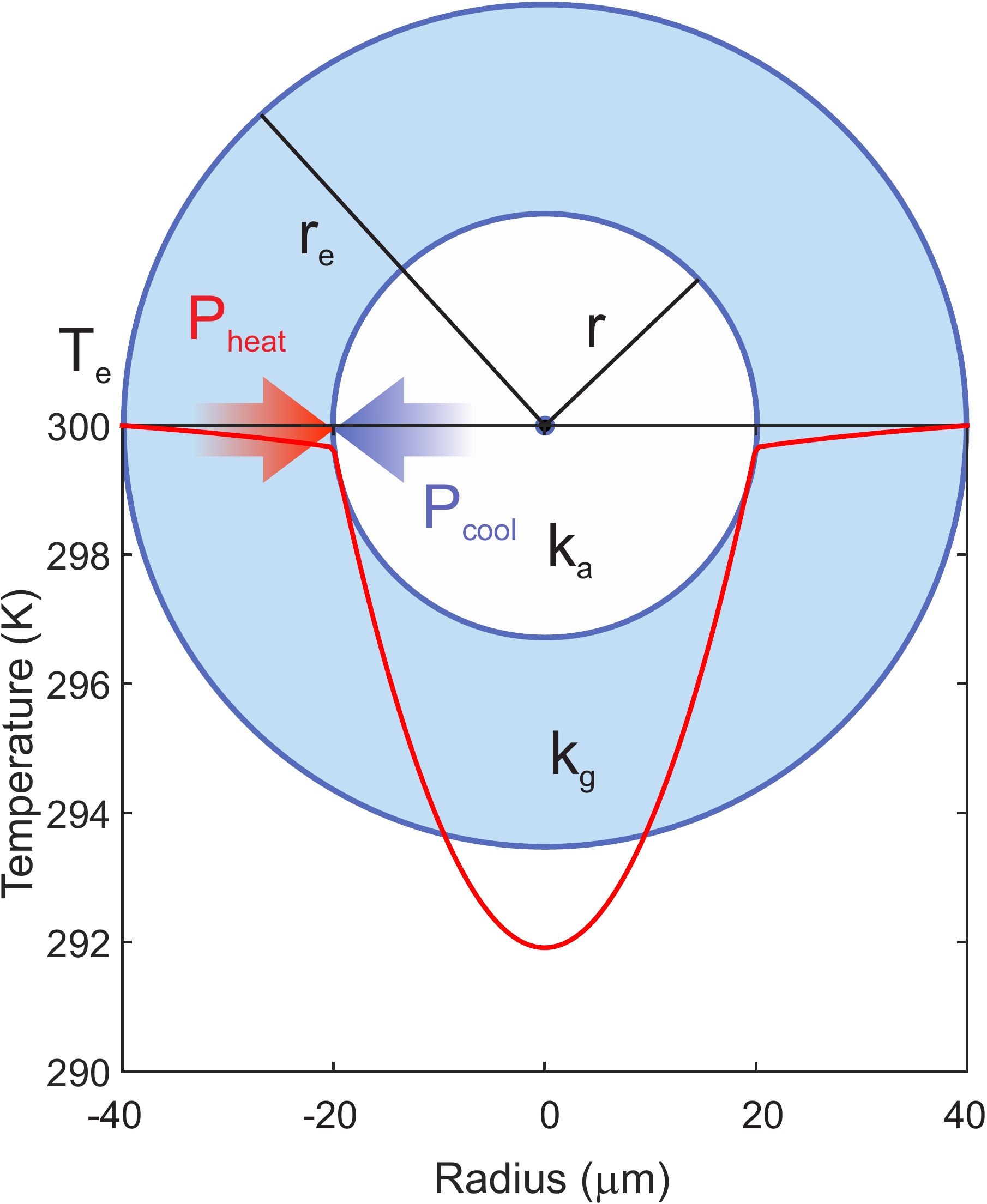}
\caption{\emph{Temperature distribution}. The plot shows the temperature distribution over the fibre cross section. The parameters to calculate the temperature distribution are taken from Table~\ref{tab1} and we have chosen the outer temperature to be $T_e = 300\,$K to display a temperature drop starting from room temperature.}
\label{fig1A}
\end{figure}
In the pumping-saturation regime where $\mathcal{P}(z) = \mathcal{P}(0) - \mathcal{A}z$ and with $\mathcal{A}z \leq \mathcal{P}(0)$ for $z \in [0,\ell]$ we can easily derive an expression for the Temperature distribution in the hollow core fibre (see Fig.~\ref{fig1A}). We have an Ar-Rb gas mixture in the fibre core starting from the radius $r$ (with a heat conductivity $k_a$ for Argon) which has a constant volumetric cooling source term
\begin{eqnarray}
\label{Ap:Temp:Eq1}
q_{vol} &=& \frac{\Omega}{\omega_{L}\pi r^2}\frac{d\mathcal{P}}{dz} = -\frac{\Omega\mathcal{A}}{\omega_{L}\pi r^2}\\\nonumber
&=& -\frac{1}{2}\hbar\Omega\gamma n_{M},
\end{eqnarray}
over the whole length $\ell$ of the fibre according to the saturation regime.
With independence of temperature on the angle and coordinate along the fibre where the latter comes from working in the saturation regime the full heat equation for this problem is given by
\begin{subequations}
  \begin{align}
	\label{Ap:Temp:Eq2}
	n_{\text{A}} c^{(A)}_{p}\partial_t T - k_a \rho^{-1}\partial_{\rho}[\rho\partial_{\rho} T] = q_{vol} \;\;\;\;\;\;\;\;\; \rho \leq r \\
	n_G c^{(G)}_{p}\partial_t T - k_g \rho^{-1}\partial_{\rho}[\rho\partial_{\rho} T] = 0 \;\;\;\; r < \rho \leq r_{e},
	\end{align}
\end{subequations}

where $n_{\text{A}} = n_{\text{X}}$ and $n_G$ here are the densities, $c^{(A)}_{p}$, $c^{(G)}_{p}$ are the specific heat capacities and $k_a$, $k_g$ the thermal conductivities for the Argon gas and the silica glass, respectively.
Since we want to obtain the steady state solution we set $\partial_t T = 0$. This results in
\begin{subequations}
  \begin{align}
	\label{Ap:Temp:Eq3}
	\partial_{\rho}[\rho\partial_{\rho} T] = -\frac{q_{vol}}{k_a}\rho \,\;\;\;\; \rho \leq r \\
	\partial_{\rho}[\rho\partial_{\rho} T] = 0 \;\;\;\;\; r < \rho \leq r_{e},
	\end{align}
\end{subequations}
For $r < \rho \leq r_e$ we obtain a straightforward solution
\begin{eqnarray}
\label{Ap:Temp:Eq4}
T(\rho) &=& A\ln\left(\frac{\rho}{r_e}\right) + T_e.
\end{eqnarray}
For $\rho \leq r$ the solution is expressed by
\begin{eqnarray}
\label{Ap:Temp:Eq5}
T(\rho) &=& -\frac{q_{vol}}{4k_a}\rho^2 + B,
\end{eqnarray}
if we use the boundary condition $\rho\partial_{\rho}T|_{\rho = 0} = 0$ resulting from rotation symmetry.
Since we observe the heating rate through the cylindrical area with radius $r$ given by $\mathcal{P}_{heat} = 2\pi \rho \ell k_g\partial_\rho T |_{\rho = r}$ and equate it to the cooling rate $2\pi \rho \ell k_a\partial_\rho T |_{\rho = r} = -q_{vol}\pi r^2 \ell = -\mathcal{P}_{cool}$ (von Neumann boundary conditions). Here, we obtain $A = -\mathcal{P}_{cool}/(2\pi \ell k_g)$. The second condition for the boundary between silica and the gas mixture is that the glass and gas temperature is the same for $\rho = r$. This results in the equation
\begin{equation}
\label{Ap:Temp:Eq6}
-\frac{\mathcal{P}_{cool}}{2\pi \ell k_g}\ln\left(\frac{r}{r_e}\right) + T_e = -\frac{\mathcal{P}_{cool}}{4\pi r^2 \ell k_a}r^2 + B,
\end{equation}
which fixes $B$.
Finally, we obtain the solutions
\begin{subequations}
  \begin{align}
	\label{Ap:Temp:Eq6}
	\nonumber
	T(\rho) = \frac{-\mathcal{P}_{cool}}{2\pi\ell}\left[\frac{1}{2k_a}\left(\frac{\rho^2}{r^2}-1\right) + \frac{1}{k_g}\ln		\left( \frac{r}{r_e}\right) \right] + T_e \; \rho \leq r \\
	\\\nonumber
	T(\rho) = -\frac{\mathcal{P}_{cool}}{2\pi\ell}\frac{1}{k_g}\ln\left(\frac{\rho}{r_e} \right) + T_e \,\;\;\;\;\;\;\;\;\;\;\;\;\;\;\;\;\;\;\;\;\;\;\; r < \rho \leq r_e,\\
\end{align}
\end{subequations}
which are illustrated in Fig.~\ref{fig1A}. More complicated geometrical structures as presented in the main text (see Fig.~5c,d) can be handled by numerical simulations of the heat equation.

\subsection{Elastic collision cross section}
The differential cross-section for the scattering of an incoming plane wave off an isotropic potential is given by:
\begin{figure}[t]
\includegraphics[width=1.00\columnwidth]{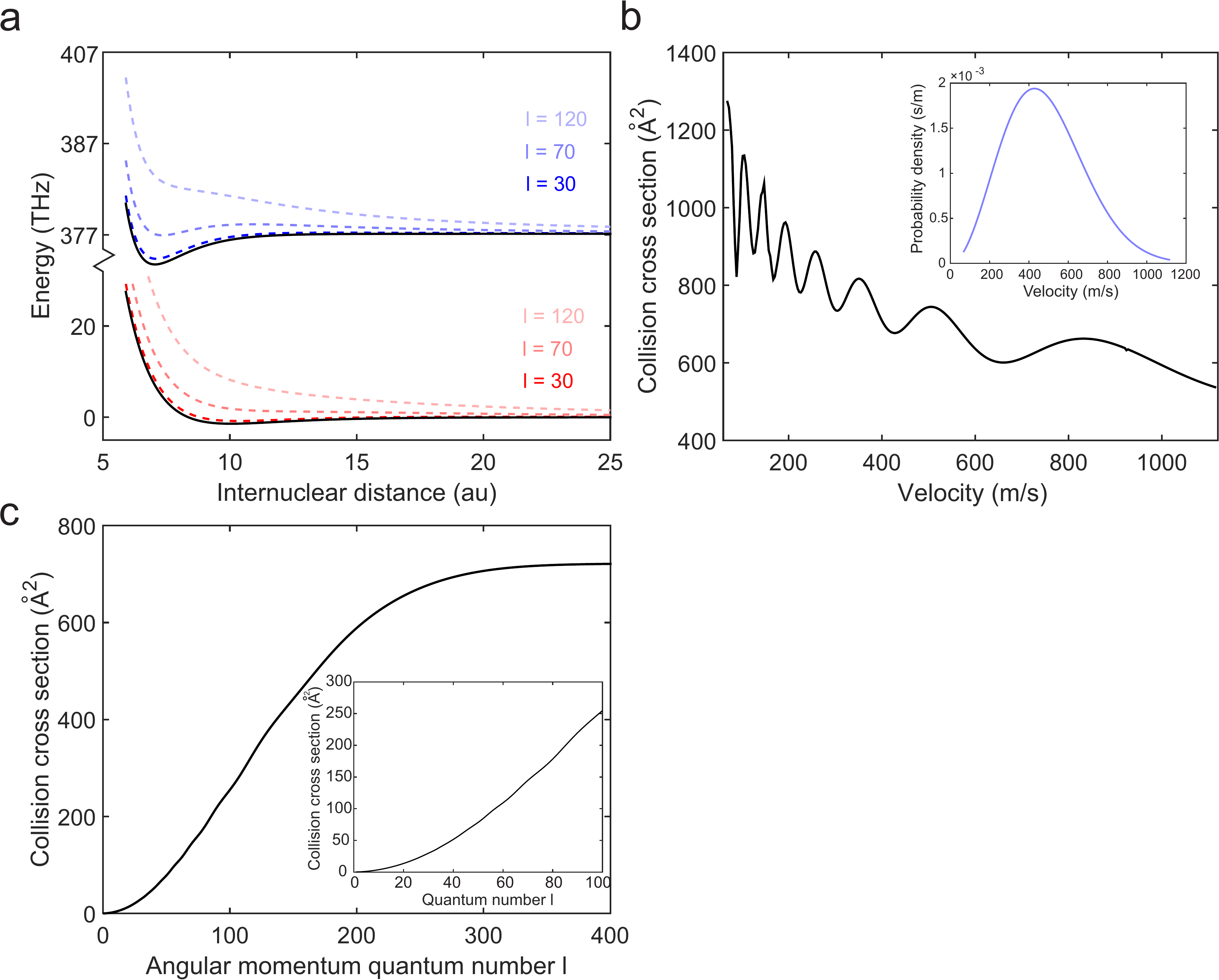}
\caption{\emph{Collision properties}. In a) the modification of the collision potentials for the ground and excited states due to the centrifugal potential is presented for three different angular momentum quantum numbers. For collisions with $l \leq 40$ the potential is not significantly modified so that the cooling process is unperturbed. b) Collision cross section for different velocities. The inset shows the velocity distribution at $T = 300\,$K. c) The collision cross section incorporating angular momenta up to $l$.}
\label{fig3A}
\end{figure}
\begin{eqnarray}
\label{Ap:Col:Eq1}
\frac{d\sigma}{d\Omega} = |f(k,\Theta)|^2,
\end{eqnarray}
where $k$ is the wavenumber and $f(k,\Theta)$ gives the scattering amplitude into outgoing spherical waves
\begin{eqnarray}
\label{Ap:Col:Eq2}
\psi = A\left( e^{ikz}+f(k,\Theta)\frac{e^{ikr}}{r}\right).
\end{eqnarray}
The scattering amplitude can be expressed by
\begin{eqnarray}
\label{Ap:Col:Eq3}
f(k,\Theta) = \frac{1}{k}\sum^{\infty}_{l=0}(2l+1)e^{i\delta_l}\sin(\delta_l)P_l(\cos(\Theta)),
\end{eqnarray}
where $l$ is the angular momentum quantum number, $P_l$ are Legendre polynomials and $\delta_l$ is the phase shift of the partial asymptotic wavefunctions due to the interaction potential.
Therefore, the total elastic cross section for a given wavenumber $k$ is
\begin{eqnarray}
\label{Ap:Col:Eq4}
\sigma(k) &=& 2\pi\int^{\pi}_{0} |f(k,\Theta)|^2 \sin(\Theta) d\Theta \\
&=& \frac{4\pi}{k^2}\sum^{\infty}_{l=0}(2l+1)\sin^2(\delta_l).
\end{eqnarray}
Using the velocity distribution
\begin{eqnarray}
\label{Ap:Col:Eq5}
P(v) = 4\pi v^2\left(\frac{\mu}{2\pi k_{\text{B}} T}\right)^{3/2}e^{-\frac{\mu v^2}{2k_{\text{B}} T}}
\end{eqnarray}
for the relative motion between the colliding species, we obtain the final cross section
\begin{eqnarray}
\label{Ap:Col:Eq6}
\sigma_{el} = \int^{\infty}_{0}dv P(v)\sigma(\mu v/\hbar).
\end{eqnarray}
For efficient cooling cycles, the centrifugal barrier contribution to the excited state potential landscape should be kept small; this in turns limits the number of $l$ values considered in the sum above for the elastic cross-section for a given $k$ wavenumber (see Fig.~\ref{fig3A}a). For the numerical simulations considered in the text we estimate an effective collisional cross section (allowing the cooling cycles) of $\sigma_{\text{cool}}$ around $20\,$\AA$^2$.

\subsection{Comparison with experimental results}
We test our theoretical model by simulating the temperature drop measured and presented in~\cite{Weitz2009laser}.
\begin{figure}[b]
\includegraphics[width=1.00\columnwidth]{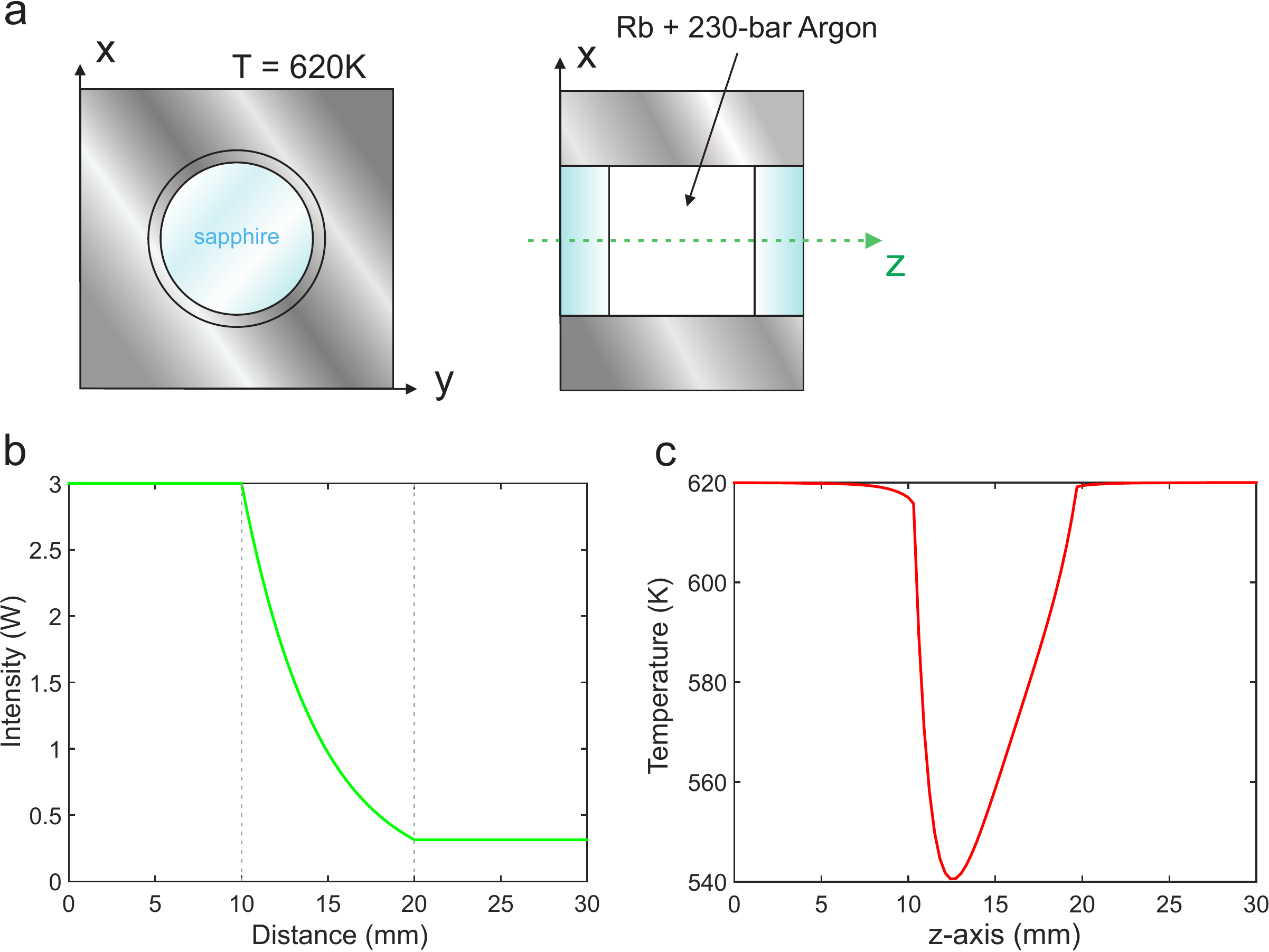}
\caption{\emph{Simulation results}. Simulation using the experimental constrains of~\cite{Weitz2009laser}. a) Sketch of the experimental set-up used in~\cite{Weitz2009laser}. b) Laser power inside and outside of the buffer-gas cell. Inside the cell the power decreases exponentially. Up to $90\%$ of the initial power is absorbed by the gas mixture in the cell over a length of $1\,$cm. c) Temperature drop along the z-axis in the center of the sapphire window. The maximum temperature drop is around $\sim 80\,$K, which is in good agreement with the result obtained in~\cite{Weitz2009laser}.}
\label{fig4A}
\end{figure}
The simulation entails a buffer gas-cell as depicted in Fig.~\ref{fig4A}a with sapphire windows of thermal conductivity $k_{\mathrm{sap}} \approx 2\,$WK$^{-1}$m$^{-1}$ at the input and output forming the boundary for the gas along the z-axis.   The temperature on the outside of the cell and the sapphire window is fixed to $T = 620\,$K. We obtain the factors $\mathcal{A}$, $\mathcal{B}$ for an Argon buffer-gas pressure of $230\,$bar which translates to a density of $n_{A} \approx 10^{21}\,$cm$^{-3}$ at $T=620\,$K as well as a Rubidium density of $n_{R} \approx 10^{16}\,$cm$^{-3}$ and a cooling beam diameter of $3\,$mm allowing us to obtain the drop of the laser power in the cell as presented in Fig.~\ref{fig4A}b.
The feature of the temperature drop via laser cooling as shown in Fig.~\ref{fig4A}c is obtained from a simulation of the heat equation following a volumetric cooling input term which is governed by the population dependent power loss and can be derived by employing the relation for the laser power in the cell presented in Fig.~\ref{fig4A}b.
The resulting maximal temperature drop of $\sim80\,$K is in good agreement with the $66\pm 13\,$K presented in \cite{Weitz2009laser}.

\end{document}